\documentclass[twocolumn,showpacs,aps,prl]{revtex4}
\usepackage{graphicx}
\usepackage{dcolumn}
\usepackage{amsmath}
\usepackage{latexsym}

\begin {document}


\title {Fiber Bundle model with Highly Disordered Breaking Thresholds}
\author
{Chandreyee Roy, Sumanta Kundu and S. S. Manna}
\email{manna@bose.res.in}
\affiliation
{
\begin {tabular}{c}
Satyendra Nath Bose National Centre for Basic Sciences,
Block-JD, Sector-III, Salt Lake, Kolkata-700098, India
\end{tabular}
}
\begin{abstract}
      We present a study of the fiber bundle model using equal load sharing dynamics 
   where the breaking thresholds of the fibers are drawn randomly from a power law distribution of 
   the form $p(b)\sim b^{-1}$ in the range $10^{-\beta}$ to $10^{\beta}$. Tuning the value
   of $\beta$ continuously over a wide range, the critical behavior of the fiber bundle has been 
   studied both analytically as well as numerically. Our results are: 
   (i) The critical load $\sigma_c(\beta,N)$ for the bundle of size $N$ approaches its asymptotic 
   value $\sigma_c(\beta)$ as $\sigma_c(\beta,N) = \sigma_c(\beta)+AN^{-1/\nu(\beta)}$ where 
   $\sigma_c(\beta)$ has been obtained analytically as $\sigma_c(\beta) = 10^\beta/(2\beta e\ln10)$ 
   for $\beta \geq \beta_u = 1/(2\ln10)$, and for $\beta<\beta_u$ the weakest fiber failure leads 
   to the catastrophic breakdown of the entire fiber bundle, similar to brittle materials, leading 
   to $\sigma_c(\beta) = 10^{-\beta}$; (ii) the fraction of broken fibers right before the complete 
   breakdown of the bundle has the form $1-1/(2\beta \ln10)$; (iii) the distribution $D(\Delta)$ of 
   the avalanches of size $\Delta$ follows a power law $D(\Delta)\sim \Delta^{-\xi}$ with $\xi = 5/2$ 
   for $\Delta \gg \Delta_c(\beta)$ and $\xi = 3/2$ for $\Delta \ll \Delta_c(\beta)$, where the 
   crossover avalanche size $\Delta_c(\beta) = 2/(1-e10^{-2\beta})^2$. 
\end{abstract}

\maketitle

\section {1. Introduction}

      Natural disasters like land slide, mine collapse, earthquake cause great losses in human 
   lives and property. It is therefore primarily important to understand the underlying mechanisms
   of the failure processes so that the losses can be minimized by providing a precursor. Similarly 
   for engineers the strength of material is a major quantity in order to make huge constructions
   like bridges, buildings etc. Due to these standing requirements, during the last two decades, 
   huge amounts of scientific efforts have been invested to explore the microscopic mechanism and 
   rupture process of disordered materials. It has been revealed that the disorder plays a crucial 
   role in determining the strength of material and also in the fracturing process \cite {Herrmann,
   Chakrabarti,Sornette,Sahimi,Bhattacharya}. 

      Models of materials in the form of a bundle consisting of a large number of parallel massless 
   elastic fibers are well known to be simple examples of critical systems exhibiting non-trivial 
   breakdown properties \cite {Herrmann, Chakrabarti,Sornette,Sahimi,Bhattacharya}. These systems 
   are called the Fiber Bundle Models (FBM) where individual fibers have randomly distributed breaking 
   thresholds. Typically, on increasing the externally applied load $\sigma$ per fiber, the entire 
   fiber bundle fails at a critical load $\sigma_c$ per fiber. It is also known that for $\sigma < 
   \sigma_c$, larger the external load, more extensive is the response of the system in terms of 
   the number of fiber failures. This number diverges as $\sigma \to \sigma_c$ from below and for 
   $\sigma$ beyond $\sigma_c$ all fibers eventually fail with certainty. Therefore, $\sigma_c$ is looked upon 
   as the transition point from a local to the global failure of the bundle \cite {Pradhan1}.

      In the fiber bundle model a set of $N$ parallel fibers is clamped at one end and an external 
   load is applied at the other end \cite {Pierce,Daniels}. Every fiber $i$ has its own breaking 
   threshold $b_i$. If the tensile stress acting through it exceeds $b_i$ it breaks. Random numbers 
   $\{b_i\}$ are drawn from a probability distribution $p(b)$ and they are assigned as the breaking 
   thresholds of the individual fibers whose cumulative distribution is $P(b) = \int^{b}_0 p(z)dz$.
    
      In many FBMs, stress is treated as a conserved quantity. During the failure of an individual fiber
   the stress is released and it gets distributed among the remaining intact fibers. Depending on how
   the released stress is distributed among the intact fibers there exists various models in the
   literature. Among these FBMs the Equal Load Sharing (ELS) model is the most well known \cite {Daniels,Andersen,Kloster}. 
   Here the released stress is distributed equally among all the remaining intact fibers. Most of the 
   results of this model have been calculated analytically and also this model is computationally easier
   to tackle. On the other hand in the Local Load Sharing (LLS) model, the released stress is 
   distributed equally only to the nearest surviving neighbors \cite {Harlow1,Harlow2}. In the LLS model, most of the
   results have been obtained numerically. A fiber is strained when some amount of stress acts through
   it. For ELS, the clamps at two ends of the bundle may be treated as infinitely stiff and therefore under 
   a certain applied load, all fibers are strained by equal amounts and consequently the magnitudes
   of the stresses acting through the intact fibers are also equal. On the other hand, if the clamps 
   are elastic, different fibers are strained differently and their stress values are also different
   as is the situation in the LLS model.
   A third model, intermediate between ELS and LLS, has also been considered in the following way. 
   In this model the released stress is distributed non-uniformly and the share amount 
   received by an intact fiber depends inversely to some power of the distance of separation from the 
   broken fiber \cite {Hidalgo}. A number of other processes have been studied in the framework of
   fiber bundle models. For example, how the damage evolves due to an environmentally assisted aging process in a fiber
   bundle model has been studied in \cite {Kun1}. In this paper, we study the breakdown properties of the fiber bundles with ELS dynamics.
   
      Let $\sigma$ be the uniform applied load per fiber initially when all fibers are intact. The 
   total amount of external load is then $F = N\sigma$. This externally applied load gets distributed
   within the bundle in a series of $T$ successive time steps. Let us denote $x_t$ as the stress per 
   intact fiber after $t$-th relaxation step. Since more and more fibers break, the stresses acting 
   through the remaining intact fibers increase. When $\sigma$ is the applied load per fiber, all fibers
   with $b_i < \sigma$ break. This stress is now distributed to $N[1 - P(\sigma)]$ intact fibers on 
   the average. After the first step if $x_1$ is the stress per fiber then $F = Nx_1[1 - P(\sigma)]$. 
   Consideration of the same mechanism in successive steps one can write:
\begin {equation}
   F = Nx_1[1 - P(\sigma)] = Nx_2[1 - P(x_1)] = Nx_3[1 - P(x_2)] ..
\end {equation}
   This process terminates after $T$ steps when the amount of stress released is not sufficient to 
   create further failure of fibers.

      If $x$ is the applied load per intact fiber in the stable state, then one can write the 
   external load $F(x)$ as a function of $x$, which is $F(x) = Nx[1 - P(x)]$ \cite {Pradhan1,Pradhan2}. 
   For a specific value of $x = x_c, F(x)$ is maximum which suggests the condition:
   $1 - P(x_c) - x_cp(x_c) = 0$. For example, for a uniform distribution of breaking thresholds one
   gets $\sigma_c = F_c/N = 1/4$ \cite {Pradhan2}.

      The failure properties of materials are highly dependent on the extent of disorder inherent in them.
   In the FBMs, this disorder appears in the breaking thresholds of the individual fibers. In this
   regard, the power law distribution of breaking thresholds is an extreme case of heterogeneous 
   disorder, where a large number of fibers have very small breaking thresholds, their numbers 
   decreases as breaking thresholds are increased, leading to few fibers with large breaking thresholds.
   It is already known in the literature that the probability of getting a warning of imminent
   breakdown of the system is higher when the material is more heterogeneous \cite {Moreira}.
   Such cases of extremely heterogeneous disorder has not been very well studied in the literature of fiber
   bundle model. Another form of strong heterogeneity has been studied where a fraction of fibers
   are completely unbreakable and the breaking thresholds of the rest are drawn from some distribution
   \cite {Hidalgo2,Hidalgo3}. In this paper we therefore address the problem of FBMs with highly heterogeneous
   power law distributed breaking thresholds of individual fibers. As a first step we study the simpler 
   problem of ELS dynamics in this model, study of the LLS version will be taken up in a future
   publication. 
   
      FBMs have a wide variety of applications. It is a versatile tool to understand conceptually 
   the underlying microscopic mechanism of fatigue \cite {Curtin}, failure of composite materials \cite {Kun}, 
   landslides \cite {Cohen} etc. Moreover, the ELS version of the FBM studied here may be used to
   study the traffic jams in roads \cite {BKC}. The traffic flow capacities of the roads can be mapped 
   to the breaking thresholds of individual fibers. The highly disorder flow rates may occur in a traffic 
   network with few highways and a large number of narrow roads connecting the highways.
   
      In section 2 we describe our study of the fiber bundle model with power law distributed breaking thresholds. We
   also describe different analytically obtained results characterizing this bundle and their numerical 
   supports. In section 3 the statistics of avalanche size distribution have been described. We
   summarize in section 4.
\begin{figure}[t]
\begin {center}
\includegraphics[width=7.0cm]{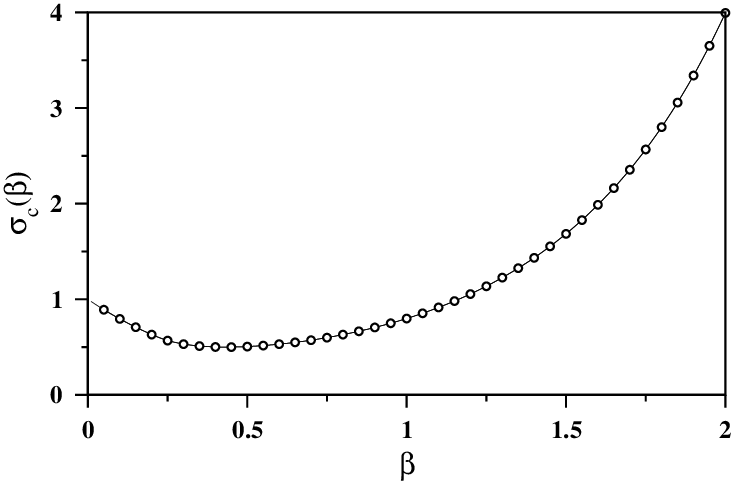}
\end {center}
\caption{The initial external load $\sigma_c(\beta)$ per fiber given in Eqn. 5 (solid line) matches excellently with its numerical
estimates (open circles).}
\end{figure}

\section {2. Highly Disordered Fiber Bundles}

      In this paper we report the results of our study of the breakdown properties of a fiber bundle where the breaking 
   thresholds of the individual fibers are power law distributed. As in other FBMs, the only source of disorder in our 
   model is the random distribution of breaking thresholds. Therefore, the individual breaking thresholds $b_i$ are drawn from a 
   probability distribution $p(b) \sim b^{-\gamma}$ with $\gamma = 1$. Initially $N$ uniformly distributed random numbers 
   $q_i$ are drawn within $-1 < q_i < 1$ and the breaking threshold $b_i = 10^{\beta q_i}$for the $i$-th fiber is assigned. 
   Consequently, the probability distribution takes the form $p(b) \sim b^{-1}$ within the range $10^{-\beta}$ to $10^{\beta}$ 
   \cite {Moreira}.

      Here, we use the same formulation described in section 1 to obtain the breaking strength of the bundle $\sigma_c$
   as a function of the cut-off parameter $\beta$ when the breaking thresholds $\{b_i\}$ are power law distributed.
   The constant of proportionality can be evaluated from the normalization condition, which gives the functional form
   $p(b) = b^{-1} / (2\beta \ln10)$. As a result, the cumulative probability distribution is given by,
\begin{equation}
   P(b) = \int_{10^{-\beta}}^b p(z)dz = \ln b / (2\beta \ln 10)+ 1 / 2.
\end{equation}
   In this case we obtain the expression of $F(x)$ as
\begin{equation}
   F(x) =  Nx [1 / 2 - \ln x / (2\beta \ln10)]. 
\end{equation}
   Clearly the function $F(x)$ has a maximum at $x=x_c$ for which $dF(x)/dx=0$. This yields $x_c=10^{\beta}/e$ and the total
   critical applied load is $F_c\equiv F(x_c)=N10^{\beta}/(2\beta e\ln10)$. Thus the critical initial applied load per fiber is
   given by
\begin{equation}
   \sigma_c(\beta) = F_c / N = 10^{\beta} /(2\beta e\ln10).
\end{equation}   
   Let $b^*$ denote the minimum of the breaking thresholds. Since the definition of $x_c$ signifies that a bundle fails completely 
   at this point then, the condition $b^* = x_c$ i.e., $10^{-\beta} = 10^{\beta}/e$ fixes the upper 
   bound of $\beta$ denoted as $\beta_u=1/(2\ln10)$ for which the weakest fiber failure leads to the complete breakdown of the bundle.  
   Thus we have the complete expression for $\sigma_c(\beta)$:

\begin{equation}
   \sigma_c(\beta) =
\left\{
\begin{aligned}
   & 10^{\beta} / (2e\ln10^{\beta})  &\quad \text{for}~ \beta \geq \beta_u \\
   & 10^{-\beta}  &\quad \text{for}~ \beta \leq \beta_u 
\end{aligned}
\right.
\end{equation}
   The above expression for the average critical applied load per fiber $\sigma_c(\beta)$ for a given value of cut-off parameter 
   $\beta$ is valid only for infinitely large bundles, i.e., $N \to \infty$. 

\begin{figure}[t]
\begin {center}
\includegraphics[width=7.0cm]{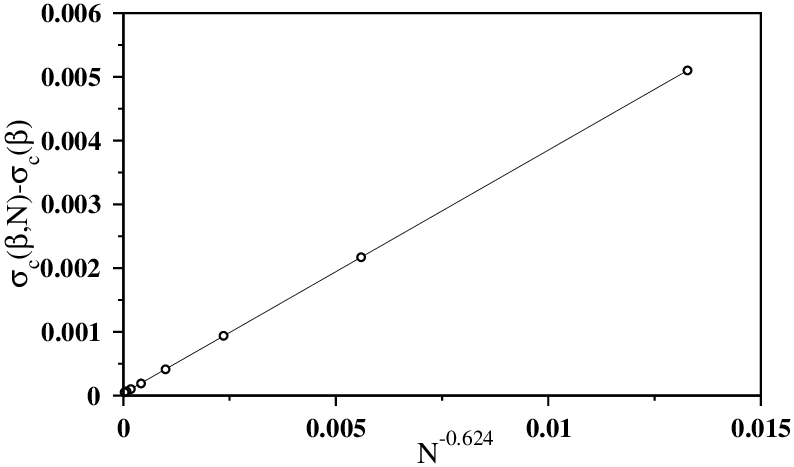}
\end {center}
\caption{Variation of the critical load $\sigma_c(\beta,N)$ on the system size $N$ for $\beta=0.225$ has been
exhibited. Plot of $\sigma_c(\beta,N)-\sigma_c(\beta)$ vs. $N^{-0.624}$ with $\sigma_c(\beta)=0.596$ shows a nice 
straight line that passes very close to the origin.} 
\end{figure}
   
         The width of the distribution of breaking thresholds increases with $\beta$ and the critical threshold $\sigma_c(\beta)$ 
   varies accordingly. For $\beta=0$, all fibers have the same breaking thresholds equal to unity and therefore $\sigma_c(0)=1$. When 
   $\beta$ is small, the minimum breaking threshold is high enough, and very close to unity. When the external stress per fiber
   is raised to reach the minimum breaking threshold, it breaks. The released stress is distributed among the remaining fibers
   and is sufficient to break all other fibers. This mechanism, when failure of the weakest fiber ensures the global failure
   of the entire bundle is analogous to the brittle fracture. This situation continues till $\beta$ reaches $\beta_u$ and therefore
   $\sigma_c(\beta)$ decreases as the strength of the weakest fiber, i.e., $10^{-\beta}$. When $\beta$ increases further, 
   gradually fibers of high breaking thresholds appear and they take over the control. Consequently, $\sigma_c(\beta)$ must
   increase with $\beta$ for large $\beta$ with a minimum at $\beta = \beta_m$. The value of $\beta_m$ is obtained using 
   the condition $d\sigma_c(\beta)/d\beta = 0$ in Eqn. (5) at $\beta_m = 1/(\ln 10)$, which is twice the value of $\beta_u$.

\begin{figure}[t]
\begin {center}
\includegraphics[width=7.0cm]{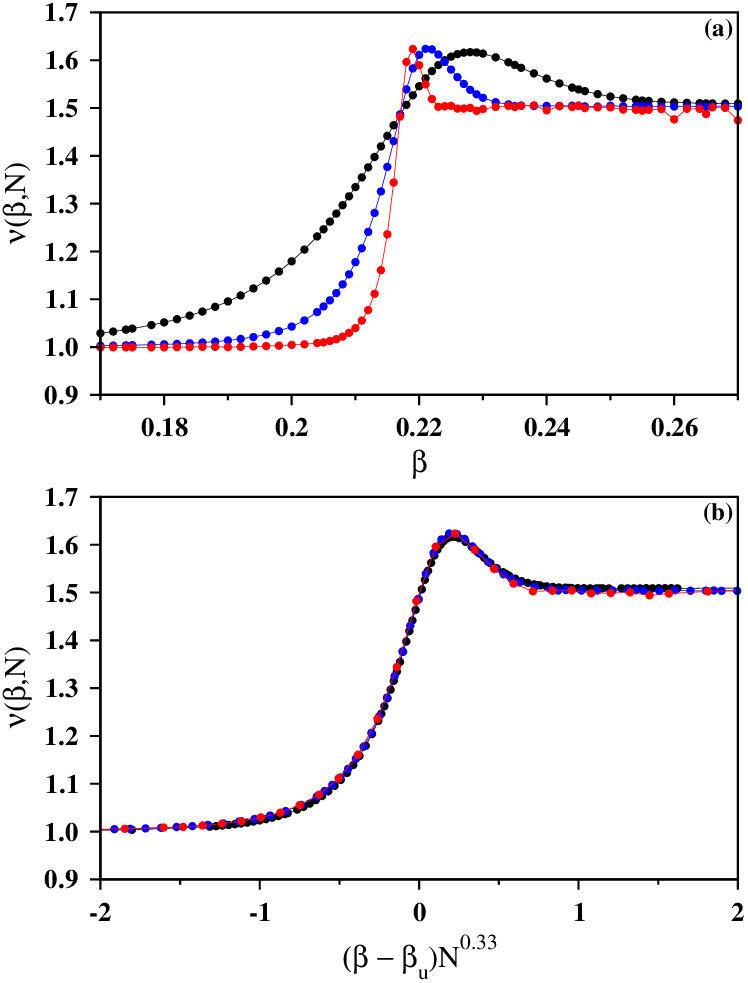}
\end {center}
\caption{(Color online) (a) Plot of $\nu(\beta,N)$ vs $\beta$ for systems of different sizes. The value of $\nu(\beta,N)$ 
calculated using the four bundle sizes from $N=2^{10}$ to $2^{16}$ (black), $2^{14}$ to $2^{20}$ (blue) and $2^{18}$ to $2^{24}$ 
(red); $N$ is increased from left to right. (b) A collapse of the data of the same three system sizes works excellent when 
the $\beta$ axis has been suitably scaled.}
\end{figure}
   
      Numerically $\sigma_c(\beta)$ is obtained in the following way. For a given value of $\beta$ we first calculate the critical 
   load per fiber $\sigma_c^\alpha(\beta,N)$ for a particular fiber bundle $\alpha$ having $N$ fibers with a given set of 
   breaking thresholds $\{b_i\}$. This calculation is repeated over a large number of un-correlated bundles $\alpha$ and their critical 
   loads are averaged to obtain $\sigma_c(\beta,N) = \langle \sigma_c^\alpha(\beta,N) \rangle$. The entire calculation is then repeated 
   for different values of $N$.

      To obtain $\sigma_c^\alpha(\beta,N)$, the breaking thresholds are arranged in increasing order ($b_{(1)}^\alpha < b_{(2)}^\alpha 
   < b_{(3)}^\alpha < ... < b_{(N)}^\alpha$). The bundle will support the initially applied load per fiber ($\sigma$) if $\sigma < b_{(1)}^\alpha$ 
   or $\sigma N/(N-1) < b_{(2)}^\alpha$ or $\sigma N/(N-2) < b_{(3)}^\alpha$ or ... $\sigma N < b_{(N)}^\alpha$. If all these inequalities 
   fail to satisfy then the bundle will no longer support the load, it will break apart. Now if $\sigma$ is such that it is sufficient to break 
   $n$ fibers, then at this stage the bundle will support the load if $\sigma N/(N-n) < b_{(n+1)}^\alpha$ i.e.,
\begin{equation}
   \sigma < [(N-n)/N]b_{(n+1)}^{\alpha}.
\end{equation}
   The term in the parenthesis of Eqn. (6) decreases with $n$ and $b^{\alpha}_{(n+1)}$ is an increasing function of $n$ as thresholds 
   are arranged in increasing order. So, the function at the right hand side of Eqn. (6) has a maximum at some $n$ and if the $\sigma$ is 
   raised at this maximum value, the bundle will break immediately. So the maximum of [$(N-n)/N$]$b_{(n+1)}^{\alpha}$ determines the critical 
   load per fiber for the bundle $\alpha$. Therefore \cite {Smith1981},
\begin{equation}
   \sigma_c^\alpha(\beta,N) = \max\Big\lbrace b_{(1)}^\alpha, \frac{N-1}{N}b_{(2)}^\alpha, \frac{N-2}{N}b_{(3)}^\alpha,...,\frac{1}{N}b_{(N)}^\alpha \Big\rbrace
\end{equation}

\begin{figure}[t]
\begin {center}
\includegraphics[width=7.0cm]{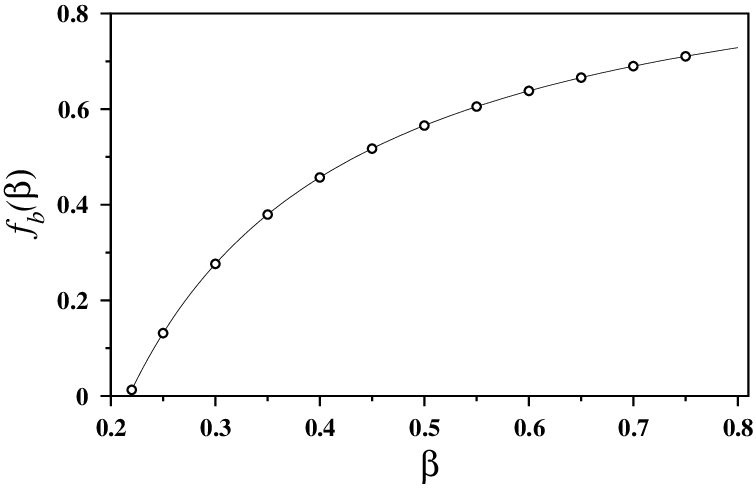}
\end {center}
\caption{Plot of the fraction of broken fibers $f_b(\beta)$ for a particular value of $\beta$ right
before the complete breakdown of the bundle is plotted for the analytical expression given in Eqn. 
9 with solid line. The numerically obtained data represented by open circles matches very well
with the analytical curve.}
\end{figure}

      We now assume that the average value of the critical load per fiber $\sigma_c(\beta,N)$ for a 
   given value of $\beta$ and for the bundle of size $N$ converges to a specific value $\sigma_c(\beta)$
   as $N \to \infty$ according to the following form:
\begin {equation}
   \sigma_c(\beta,N)-\sigma_c(\beta) = AN^{-1/\nu(\beta)}
\end {equation}
   where $\nu(\beta)$ is a critical exponent for the cut-off parameter $\beta$. We have plotted 
   $\sigma_c(\beta,N)$ against $N^{-1/\nu(\beta)}$ for $N = 2^{18}$ to $2^{24}$, $N$ being increased 
   by a factor of 4 at each stage. For a particular value of $\beta$ we have used different trial values 
   of $\nu(\beta)$ so that for a specific value of $\nu(\beta)$ the plot fits (by least square fit) 
   to the best straight line. Using this best value of $\nu(\beta)$ and on extrapolation to $N \to \infty$ 
   we obtained $\sigma_c(\beta)$. In Fig. 1 we have exhibited an excellent matching of the analytical and 
   the numerical values of $\sigma_c(\beta)$ for the range $0 < \beta \leq 2$.   

\begin{figure}[t]
\begin{center}
\includegraphics[width=7.0cm]{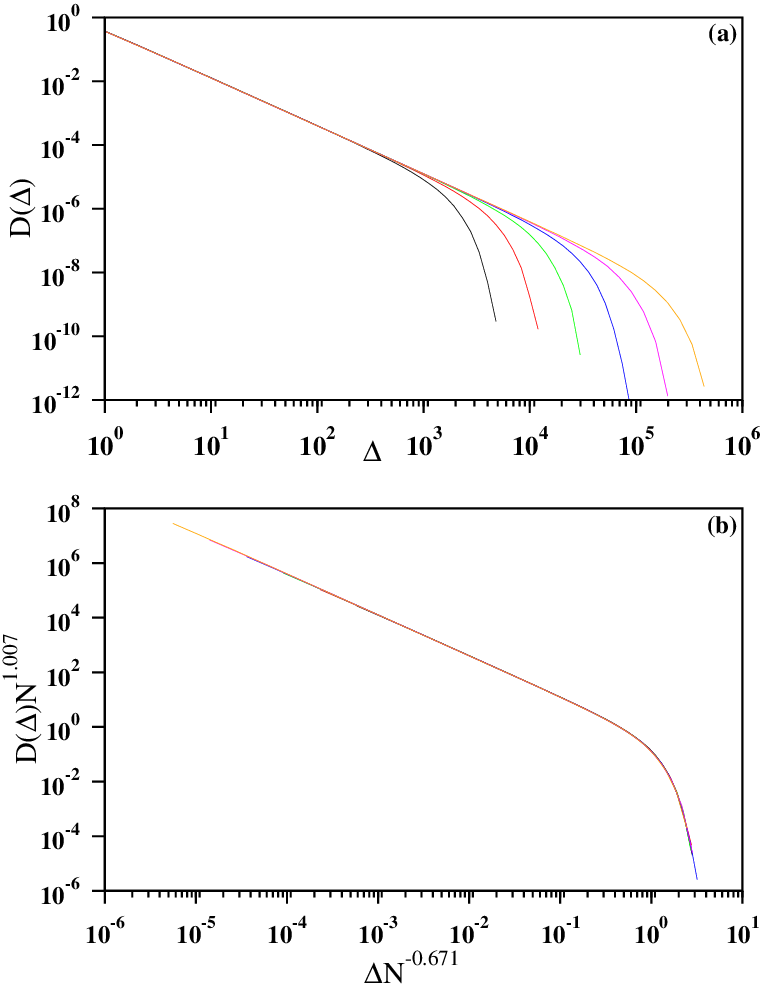}
\end{center}
\caption{(Color online) 
(a) Log-log plot of the binned data for avalanche size distribution $D(\Delta)$ vs $\Delta$ for $\beta = \beta_u = 1/(2\ln 10)$ 
    for $N= 2^{16}$, $N= 2^{18}$ ... $2^{26}$ (from left to right).
(b) A finite-size scaling works well: $D(\Delta) N^{\eta}$ against $\Delta N^{-\zeta}$ exhibits a good collapse of data with 
    $\eta = 1.007$ and $\zeta = 0.671$ implying $\xi = \eta / \zeta = 1.50(1)$. This value is consistent with the directly measured
    value of 1.50(2) from the slopes in the intermediate region. The crossover is not observed here since $\Delta_c = \infty$ 
    for this particular value of $\beta$. }
\end{figure}
   
      We now investigate the dependence of the finite size correction exponent $\nu(\beta)$ on the cut-off parameter 
   $\beta$. We recall that in the case of a uniform breaking threshold distribution, the plot of $\sigma_c(N) - \sigma_c$ 
   as a function of $N^{-1/\nu}$ gives an excellent straight line with $\sigma_c = 1/4$ and $\nu = 3/2$ 
   \cite {Roy,Smith1982,McCartney1983,Daniels2}. Similarly, for our model of highly disordered FBM, the plot of $\sigma_c(\beta,N) - 
   \sigma_c(\beta)$ against $N^{-1/\nu(\beta)}$ is carried out for different values of $\beta$. For example, we obtain 
   the best possible value of $\nu(\beta)$ to be $1.603$ for $\beta = 0.225$ shown in Fig. 2. In this way the critical 
   exponent $\nu(\beta)$ is calculated for different $\beta$ and its variation is shown in Fig. 3(a) using $N = 2^{10}$
   to $2^{16}$, $2^{14}$ to $2^{20}$ and $2^{18}$ to $2^{24}$. The value of $\nu(\beta)$ first increases,
   attains a maximum value $\approx 1.63$, then decreases and saturates to $1.5$ with further increment of $\beta$. 
   The same data in Fig. 3(a) when plotted against $(\beta - \beta_u)N^{0.33}$ shows a good collapse as shown in Fig. 3(b). 
   Thus we conclude that the curve for $\nu(\beta)$ retains its nature for large bundle sizes.
      
        Next we calculate the fraction of broken fibers $f_b(\beta)$ just before complete breakdown of the bundle as a function of the 
   cut-off parameter $\beta$. Since at $x_c$ the fiber bundle fails completely, so the quantity $f_b(\beta)$ is calculated as:
\begin{equation}
   f_b(\beta) = \int_{10^{-\beta}}^{x_c}p(x)dx = 1-1 / (2\beta \ln10). 
\end{equation}
   As the fraction of broken fibers $f_b(\beta)$ is a positive quantity thus the condition $1-1/(2\beta\ln10)>0$ again reproduces the result that
   for $\beta<1/(2\ln10)$ the weakest element failure leads to the catastrophic breakdown of the bundle. 

\begin{figure}[t]
\begin{center}
\includegraphics[width=7.0cm]{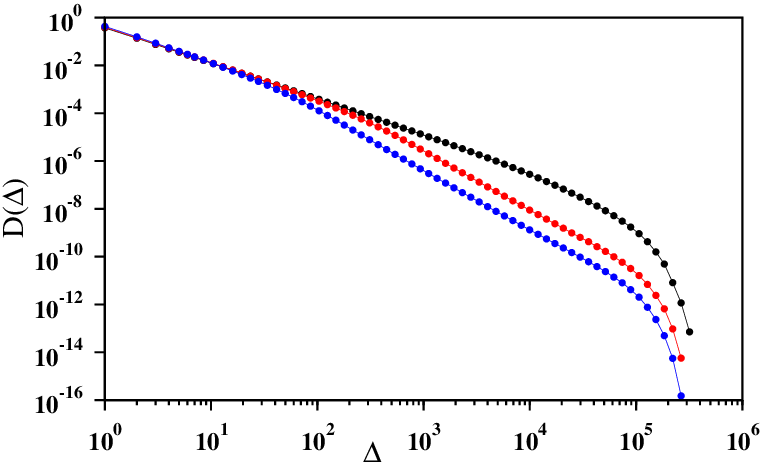}
\end{center}
\caption{(Color online) The avalanche size distribution for $\beta$ = 0.22 (black), 0.24 (red) and 0.28 (blue) (from right
to left) for bundles of size $N=2^{24}$. Slopes of the curve are $\approx 1.5$ and $\approx 2.5$ for small and large 
avalanche sizes. The crossover size $\Delta_c(\beta)$ are approximately 11741, 200.4 and 31.66 respectively evaluated using Eqn. 17.}
\end{figure}
   
      Numerically $f_b(\beta)$, for a given value of $\beta$ is calculated in the same way as described previously in the case of $\sigma_c(\beta)$.
   The external load is increased quasi-statically until the bundle fails. Just before complete breakdown of the bundle, the fraction
   of broken fibers is calculated for a particular $N$ and averaged over large number of samples to obtain $f_b(\beta,N)$. Then this procedure
   is repeated for six values of $N = 2^{16}, 2^{18}, ..., 2^{26}$ and an extrapolation on $f_b(\beta,N)$ as $N \to \infty$ yields 
   $f_b(\beta)$. In Fig. 4 the numerically obtained results are compared with the analytical one, indicating a good agreement. 

\section {3. Avalanche Size Distribution}   
   
      In a stable fiber bundle, the stress acting through every intact fiber is less than its breaking threshold.
   Now, if the externally applied load is suitably raised so that it becomes equal to the breaking threshold of
   the weakest fiber, then this fiber breaks. This triggers a cascade of fiber failures which finally ends when
   the bundle attains a new stable state. The total number $\Delta$ of fibers that fail in this event is called
   the avalanche size. Starting from a completely intact fiber bundle the global failure of the entire bundle may
   be attained by raising the external load in such a quasi-static process, causing a series of avalanches. The 
   probability distribution $D(\Delta)$ is regarded as an interesting quantity to study. It is well known that for
   the uniform distribution of breaking thresholds of individual fibers and for the ELS dynamics, the probability 
   distribution is a power law \cite {Hemmer}
\begin{equation}
   D(\Delta) \sim \Delta^{-\xi},
\end{equation} 
   with $\xi=5/2$. In the following we would see that in our case of power law distributed breaking thresholds
   the exponent $\xi$ undergoes a crossover from 3/2 to 5/2.

      To exhibit the crossover behavior we follow the method in \cite {Pradhan3}. For a bundle having large number
   of fibers, the number of avalanches of size $\Delta$ is given by \cite {Hemmer}
\begin{equation}
   \frac{D(\Delta)}{N} = \frac{\Delta^{\Delta-1}e^{-\Delta}}{\Delta!}\int_0^{x_c}p(x)r(x)[1-r(x)]^{\Delta-1}e^{\Delta r(x)}dx,
\end{equation}
   where, 
\begin{equation}
   r(x) = 1-\frac{xp(x)}{1-P(x)}. 
\end{equation}
   The expression for $D(\Delta)$ can be simplified to the following form \cite {Pradhan3}:
\begin{equation}
   \frac{D(\Delta)}{N} = \frac{\Delta^{\Delta-2}e^{-\Delta}}{\Delta!}\frac{p(x_c)}{|r'(x_c)|}(1-e^{-\Delta/\Delta_c}),
\end{equation}
   with
\begin{equation}
   \Delta_c = \frac{2}{r'(x_c)^2(x_c-b^*)^2}. 
\end{equation}
   Using the Stirling approximation $\Delta ! = \Delta^\Delta e^{-\Delta} \sqrt{2\pi \Delta}$, Eqn. (13) can be written as 
\begin{equation}
   \frac{D(\Delta)}{N} = C\Delta^{-5/2}(1-e^{-\Delta/\Delta_c}), 
\end{equation}
   Where $C = (2\pi)^{-1/2}p(x_c)/|r'(x_c)|$ is a constant. From Eqn. (15), a clear evidence of crossover in the 
   exponent $\xi$ around the avalanche size $\Delta_c$ is prominent. So we have:

\begin{equation}
   \frac{D(\Delta)}{N} \propto 
\left\{
\begin{aligned}
   \Delta^{-3/2}  &\quad \text{for}~ \Delta \ll \Delta_c, \\
   \Delta^{-5/2}  &\quad \text{for}~ \Delta \gg \Delta_c. 
\end{aligned}
\right.
\end{equation}

      In our case, we use power law distribution $p(b) \sim b^{-1}$ in the range from $10^{-\beta}$ to $10^{\beta}$
   to obtain $r'(x_c) = -e / 10^{\beta}$, $x_c = 10^{\beta}/e$ and $b^* = 10^{-\beta}$. Substituting these values in Eqn. (14)
   we get the crossover avalanche size:
\begin{equation}
   \Delta_c(\beta) = \frac{2}{(1-e10^{-2\beta})^2}.
\end{equation}

     This crossover phenomenon has also been studied using numerical simulations. For $\beta = 1/(2\ln10)$ Eqn. (17) yields 
   $\Delta_c = \infty$. Thus only the $\xi = 3/2$ power law is observed as any avalanche of finite size $\Delta$ is less 
   than the value of $\Delta_c$ at this particular value of $\beta$. In Fig. 5(a), the numerical data for the avalanche 
   size distribution for $\beta=1/(2\ln10)$ has been plotted for six different values of $N$ starting from $N=2^{16}$ to 
   $2^{26}$; $N$ being increased by a factor of 4 at each stage. For $N = 2^{16}$ to $2^{22}$ the data has been averaged 
   over $10^6$ samples and 400000 and 100000 samples for $2^{24}$ and $2^{26}$ respectively. A finite-size scaling has 
   also been done in Fig. 5(b) by use of suitable powers of the bundle size $N$. This indeed exhibits an excellent data 
   collapse confirming the following scaling form:
\begin {equation}
   D(\Delta) N^{\eta} \sim {\cal G}[\Delta / N^{\zeta}]
\end {equation}
   where ${\cal G}(y)$ is an universal scaling function of the scaled variable $y = \Delta / N^{\zeta}$. The best possible 
   tuned values of the scaling exponents obtained are $\eta = 1.007$ and $\zeta = 0.671$. Using these scaling exponents 
   the value of $\xi = \eta / \zeta = 1.50(1)$ is calculated, which is a very well tally with the analytical result of 3/2.

\begin{figure}[t]
\begin{center}
\includegraphics[width=7.0cm]{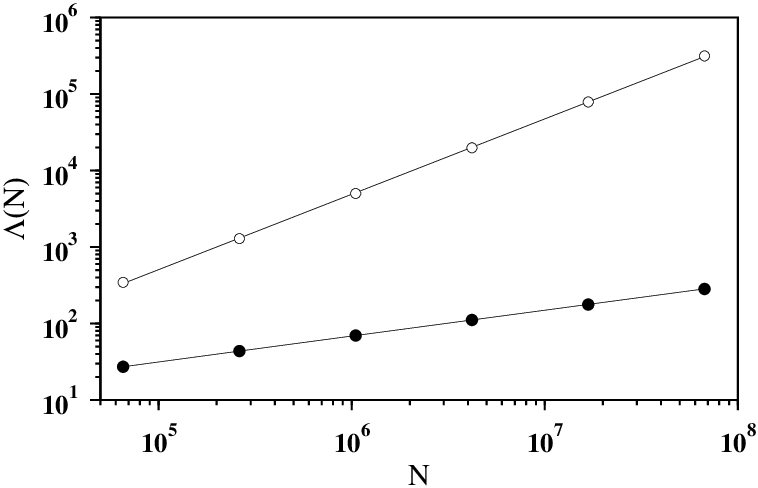}
\end{center}
\caption{Plot of the average number of avalanches $\Lambda(N)$ required to break the bundle of size $N$ on a $\log$-$\log$ scale: 
for $\beta = 1/(2\ln10)$, $\Lambda(N) \sim N^{0.337}$ (filled circles) and for $\beta = 0.240$, $\Lambda(N) \sim N^{0.985}$ (open circles). 
}
\end{figure}
   
      We have also tried the same analysis for $\beta = 0.22, 0.24$ and $0.28$. Using Eqn. (17) we have obtained $\Delta_c(\beta)
   = 11741, 200.4$ and $31.66$ respectively. A clear evidence of the crossover in the exponent $\xi$ around $\Delta = \Delta_c(\beta)$ 
   is observed as shown in Fig. 6 for $N = 2^{24}$. The slope of the curve gradually crosses over from $\approx 1.5$ to $\approx 2.5$ 
   for large values of $\Delta$. It has also been observed that as $\beta$ is increased, $\Delta_c(\beta)$ gradually shifts towards 
   the origin and therefore the regime over which $\xi = 5/2$ is valid, gets extended. Such a crossover has been observed earlier in
   \cite {Pradhan3,Pradhan4} for the FBM with uniform distribution of breaking thresholds ranged between a certain lower cutoff $b_{lc}$ and 
   unity. Here, avalanche sizes smaller (larger) than some crossover size $\Delta(b_{lc})$ correspond to avalanche size exponents 3/2 (5/2). 
   This implies that in our model, even for the highly heterogeneous distribution of breaking thresholds, similar crossover between the 
   same two exponents takes place across the crossover avalanche size $\Delta_c(\beta)$.

      It has also been observed that the total number of avalanches $\Lambda(N)$ depends on the system sizes $N$ as $N^{\chi}$, 
   where $\chi = 0.336$ and $0.985$ for $\beta = 1/(2\ln10)$ and $0.240$ respectively. The $\log-\log$ plot of $\Lambda(N)$ 
   against $N$ for these two values of $\beta$ fits to excellent straight lines as shown in Fig. 7. We conjecture that $\chi$
   may be 1/3 and 1 exactly for $\beta = \beta_u$ and $\beta > \beta_u$ respectively.
 
 \section {4. Summary}
 
      Properties of the fiber bundle model have been studied using equal load sharing dynamics where the breaking thresholds 
   of the fibers have been assigned from a power law distribution $p(b)\sim b^{-1}$ in the range from $10^{-\beta}$ to $10^{\beta}$. 
   Variations of different quantities characterizing the bundle have been studied with the cut-off exponent $\beta$. The critical 
   external load per fiber $\sigma_c(\beta)$ required for the global breakdown of the bundle as well as the fraction $f_b(\beta)$ 
   of broken fibers right before it are estimated both analytically as well as numerically, and a good correspondence has been 
   observed. For very small and very high values of $\beta$ the breaking strength of only a single fiber determines the 
   critical strength of the entire bundle. For example, for very small $\beta$, it is enough to tune the external load to the strength
   of the weakest fiber which then triggers a large avalanche and the entire bundle fails, implying $\sigma_c(\beta) = 10^{-\beta}$.
   Such a behavior continues till $\beta = \beta_u$ and this regime is analogous to the brittle failure of materials.
   When $\beta$ is raised beyond $\beta_u$, equating the external load to the strength of the weakest fiber is no more
   sufficient for the global failure, the large number of fibers with breaking thresholds near the minimum still dominate. 
   Consequently the number of avalanches required for the breakdown of the bundle gradually increases and the $\sigma_c(\beta)$
   slowly increases from the weakest strength of $10^{-\beta}$ but for $\beta > \beta_u$, $\sigma_c(\beta)$ remains smaller than
   $10^{-\beta_u}$. Therefore, a minimum in $\sigma_c(\beta)$ is reached at $\beta_m = 2\beta_u$ and from this point, $\sigma_c(\beta)$
   starts increasing. As a result, $\sigma_c(\beta)$ becomes equal to $10^{-\beta_u}$ again and then increases indefinitely. For 
   very large $\beta$ the external load must be raised to $\sigma_c(\beta) \approx 10^{\beta}$ to break the strongest fiber of breaking threshold 
   around $10^{\beta}$. This salient feature is a direct consequence of the power law distribution of the breaking thresholds. 
   
      More interestingly, we have also shown numerically that the 
   critical load $\sigma_c(\beta,N)$ approaches its asymptotic value as $\sigma_c(\beta,N) = \sigma_c(\beta)+AN^{-1/\nu(\beta)}$. The 
   finite size correction exponent $\nu(\beta)$ is first seen to increase sharply with $\beta$, reaches a maximum, then decreases and 
   finally converges to a value $\approx 3/2$. Statistical analysis of the avalanche sizes have been done. The avalanche size 
   distribution follows a power law and the associated exponent $\xi$ crosses over from $3/2$ to $5/2$ through a crossover avalanche 
   size $\Delta_c(\beta)$.  

\section{Acknowledgement}
       We thankfully acknowledge Alex Hansen for suggesting this problem and many useful discussions.

\end {document}